\documentclass{chi-ext}
\copyrightinfo{
Copyright is held by the author/owner(s).\\
{\confname{CSCW'13 Companion}}, Feb. 23--27, 2013, San Antonio, TX, USA.\\
ACM 978-1-4503-1332-2/13/02.}

\title{Truthy: Enabling the Study of Online Social Networks}

\numberofauthors{2}
\author{
  \vspace{-1.5em} 
  \alignauthor{
      \textbf{Karissa McKelvey}\\ 
  }
  \alignauthor{
      \textbf{Filippo Menczer}\\
      \affaddr{Center for Complex Networks and Systems Research}\\
      \affaddr{Indiana University}\\
      \affaddr{Bloomington, IN, USA}
  }
}

\def\plaintitle{Truthy: Enabling the Study of Online Social Networks}
\def\plainauthor{Karissa McKelvey, Filippo Menczer}
\def\plainkeywords{Research Data Management; Visualization; HCID; Collective Intelligence; Online Social Networks}
\def\plaingeneralterms{Design; Human Factors}

\hypersetup{
  pdftitle={\plaintitle},
  pdfauthor={\plainauthor},  
  pdfkeywords={\plainkeywords},
  pdfsubject={\plaingeneralterms},
}

\usepackage{graphicx}   
\usepackage{balance}    
\usepackage{bibspacing}

\begin{document}

\maketitle

\begin{abstract}
The broad adoption of online social networking platforms has made it possible to study communication networks at an unprecedented scale. Digital trace data can be compiled into large data sets of online discourse. However, it is a challenge to collect, store, filter, and analyze large amounts of data, even by experts in the computational sciences. Here we describe our recent extensions to Truthy, a system that collects Twitter data to analyze discourse in near real-time. We introduce several interactive visualizations and analytical tools with the goal of enabling citizens, journalists, and researchers to understand and study online social networks at multiple scales.

\end{abstract}

\keywords{\plainkeywords}

\category{H.5}{Human-centered computing}{Information visualization; Collaborative and social computing devices} \category{H.3.7}{Information systems}{Digital libraries and archives; Data stream mining; Collaborative filtering}. 

\section{Introduction}

Reseachers can study commnication networks on a larger scale than has been possible before in human history due to the high availability and use of the Internet as a central communication platform~\cite{vespignani}. Recent studies have demonstrated that digital trace data can be combined with sophisticated statistical tools to produce insights into the behavior and interactivity patterns of hundreds of thousands of individual actors~\cite{lazer,polarization}. With these new information and communication technologies, researchers are afforded with many opportunities to further human knowledge and understanding of community discourse and deliberation.
 
Despite the promises of these advances, there are various limitations to using social networks as a primary data source. It is often difficult or expensive for researchers trained in the social sciences to utilize the techonological expertise required to collect, store, filter, and analyze large amounts of data.  Spam and misinformation add noise to results, often originating from compromised accounts of otherwise legitimate users~\cite{spam-book-ch2,grier,politicalabuse}, and should be flagged for removal. Research is also difficult to reproduce when performed on a wide variety of datasets gathered with custom toolkits. Thus, it is beneficial for researchers to utilize a centralized platform. It is also important that any such platform be free or inexpensive, unlike for-profit social media analytics services such as GNIP (\url{http://gnip.com/}), as researchers often operate within limited budgets. 

We demonstrate the following contributions: interactive visualizations enabling users to better understand social networks and online trends; tools allowing users to freely download derived data from our large historical repository of online discourse; interfaces that allow users to tag content in order to improve automatic classifiers that detect behavior such as spam and misinformation.

\section{The Truthy System}
The \emph{Truthy} system (\url{http://truthy.indiana.edu}) was originally designed to analyze and detect the emergence of coordinated misinformation campaigns on Twitter~\cite{politicalabuse}. Now tasked with advancing the study of social networks in general, Truthy monitors a real-time feed of 140-character messages known as tweets and clusters them into groups of related messaged called ``memes.'' 

\begin{figure}
 \centering
 \includegraphics[width=\linewidth]{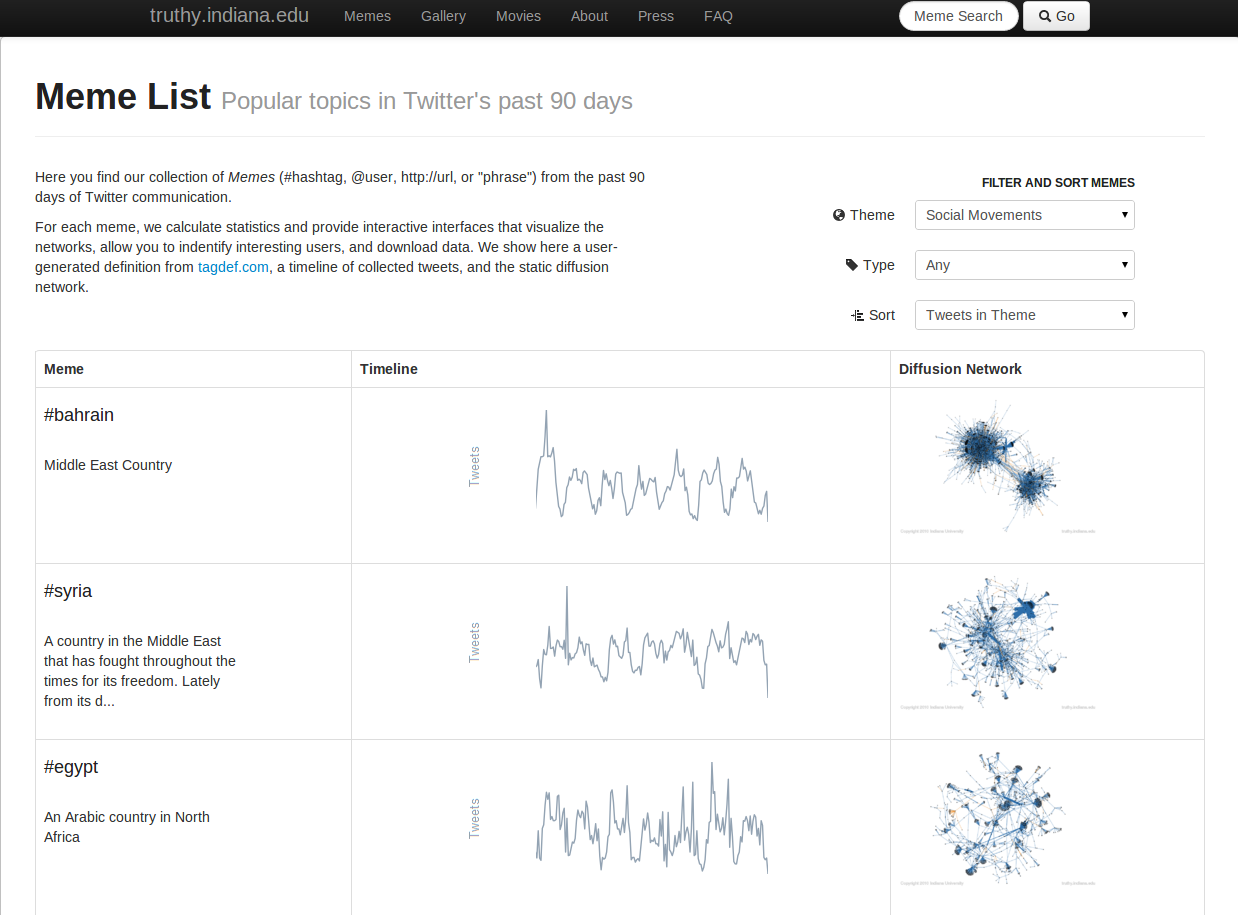}
 \caption{Theme detail page for Social Movements. One can click a meme to reach its meme detail page.}
 \label{fig:themedetail}
\end{figure}

Memes typically correspond to discussion topics, communication channels, or information resources shared among Twitter users, so that one can focus attention to understandable units of information transfer. We define each meme as the set of all tweets containing a common hashtag  (e.g., \texttt{\#bahrain}), mentioned username (e.g., \texttt{@BarackObama}), hyperlink, or phrase.  Memes are extracted from tweets that match lists of hand-picked keywords (``themes''); users can browse memes according to these themes (Fig.~\ref{fig:themedetail}). We refer the reader to previous work for a detailed description of the algorithms utilized for data 
collection, storage and filtering~\cite{Truthy_www2011demo}.

\begin{figure}
\centering
\includegraphics[width=\linewidth]{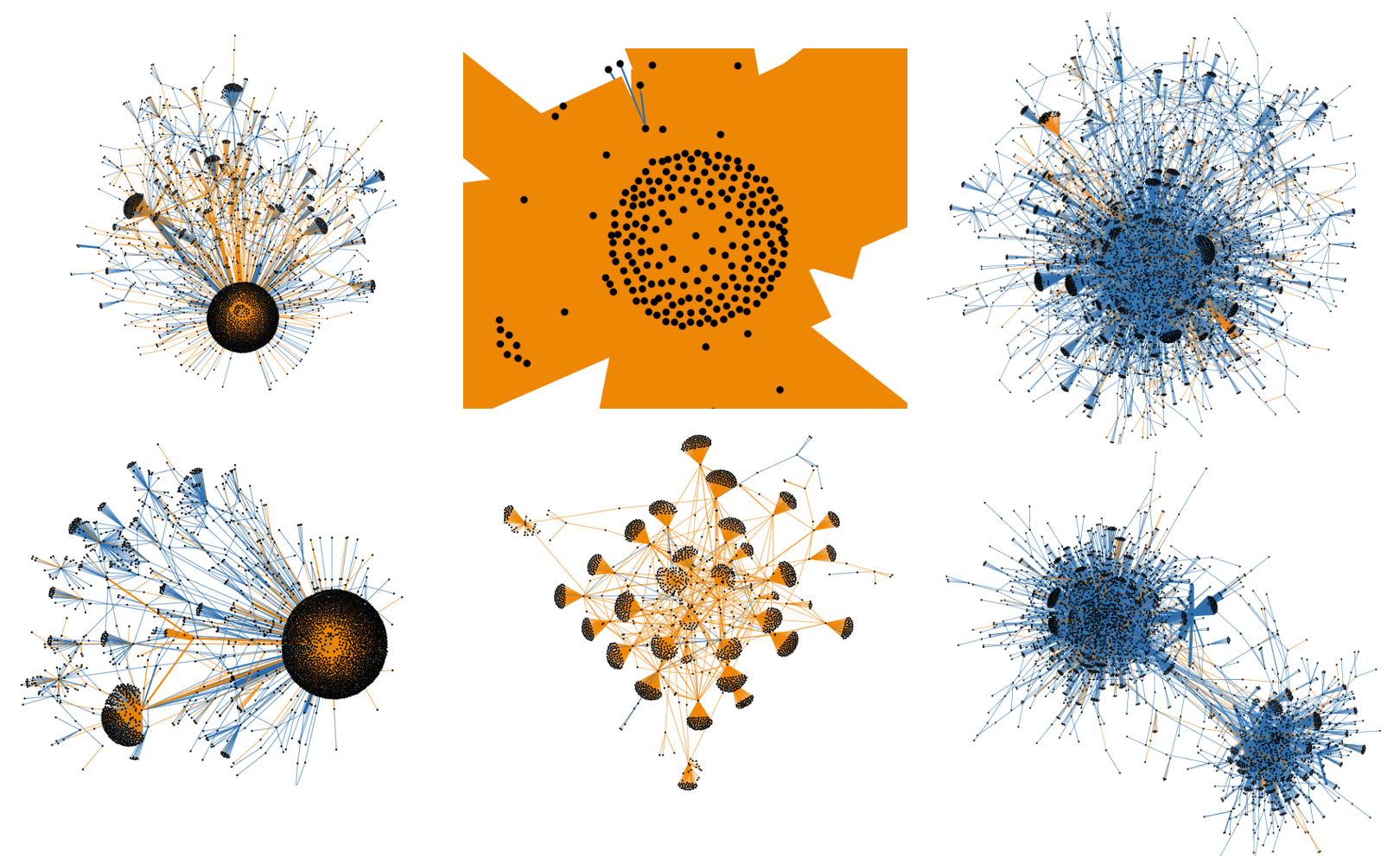}
\caption{Diffusion networks associated with multiple Twitter memes. Nodes represent individual users and edges show how these memes spread from user to user by way of mentions (orange) and retweets (blue). In clockwise order: @whitehouse, \#nightclub, \#tcot, \#syria, \#rsvp, @michelleobama}
\label{fig:diffusion}
\end{figure}

For each meme, the user is presented with an interactive dashboard containing a crowdsourced definition from Tagdef (\url{http://tagdef.com}), a high-resolution image of the meme's information diffusion network (Fig.~\ref{fig:diffusion}), and various interactive visualizations and statistics. Available information includes the number of users and tweets, meme diffusion network statistics such as mean degree and largest connected component size, as well as user-specific statistics such as predicted political partisanship, sentiment score, language, and activity.  Users can download the aforementioned derived data,  recent tweets, and the network graphs themselves for use in a spreadsheet or an analysis application such as Network Workbench~\cite{networkWorkbench}. We ensure that this data download function abides by the Twitter Terms of Service.

\begin{figure*}
 \centering
 \includegraphics[width=\textwidth]{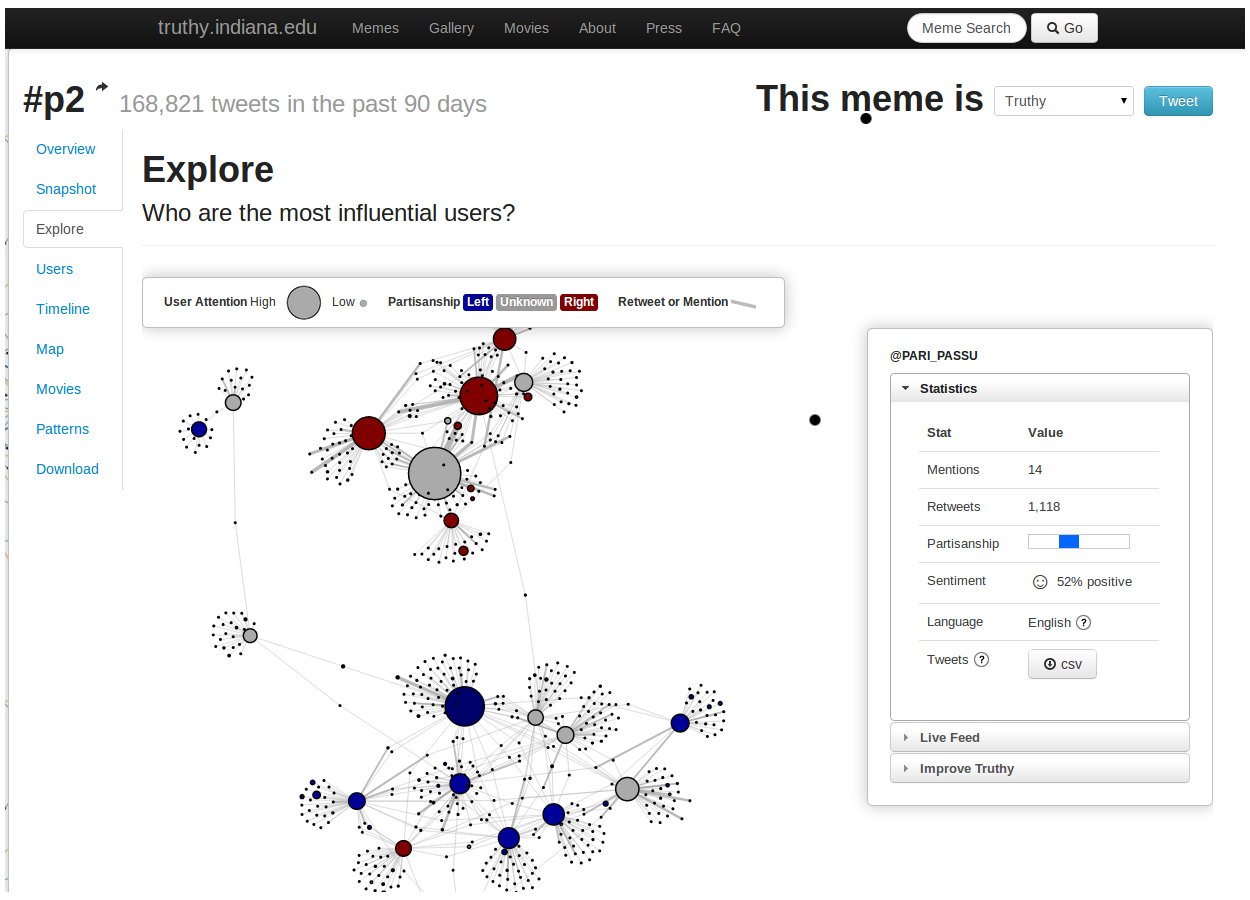}
 \caption{Meme detail page for \#p2, focusing on the interactive visualization depicting the most active users. Node size is a function of the number of retweets. Left-leaning users are colored blue while right-leaning are colored red. One can click any node in order to learn more about that user.}
 \label{fig:memedetail}
\end{figure*}

\begin{figure}
 \centering
 \includegraphics[width=\linewidth]{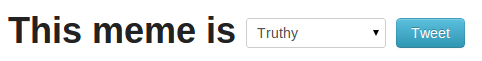}
 \caption{Push-button tool which allows one to tweet about a particular meme. Options include ``Truthy'' and ``Spam.''}
 \label{fig:truthymeme}
\end{figure}

\begin{figure}
 \centering
 \includegraphics[width=\linewidth]{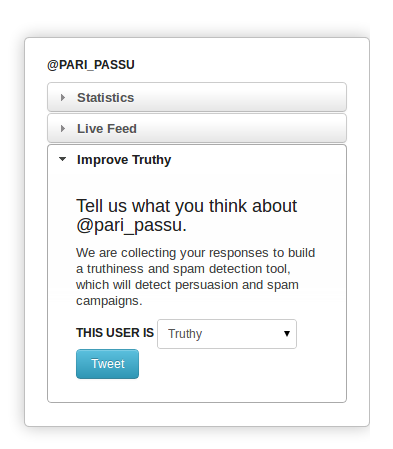}
 \caption{Interface one sees after clicking on a user in the interactive network.}
 \label{fig:truthyuser}
\end{figure}

We provide various interactive visualizations allowing users to interrogate the network structure, the characteristics of geography and time, and meme-meme co-occurrence patterns~\cite{McKelvey2012viz}. Another important new contribution of the Truthy system is in enabling active improvement of our algorithms by facilitating the tagging of suspicious content. Through the meme detail interface (Fig.~\ref{fig:memedetail}), one can tweet about a meme (Fig.~\ref{fig:truthymeme}) or user (Fig.~\ref{fig:truthyuser}) in a syntax that can be automatically parsed by our system. We collect these posts for future studies analyzing the reliability of crowdsourced data in identifying persuasion and spam campaigns. 

\section{Future Work}
In future work, we would like to provide a broader set of historical data while improving our visualizations via collaboration with the public. These initiatives include: providing a public \texttt{REST} API for derived data about individual tweets, memes, and users; expanding the scope to include all of the tweets we have collected since September 2010; facilitating user-defined themes to expand visualizations to customized content.

\section{Conclusion}
Tools like Truthy stimulate the study of online social networks. In particular, we aim to support social scientists who currently find this research difficult, expensive, or impossible to reproduce. We hope that open data and open tools like ours will advance efforts to leverage large data streams from the Internet as a primary source in the social sciences. Furthermore, we hope that our interactive visualizations facilitate the navigation and understanding of online discourse, whether for research, journalism, or general use cases.

\section{Acknowledgements}
We would like to thank Clayton Davis, Michael Conover, Jacob Ratkiewicz, Bruno Goncalves, Mark Meiss, Alessandro Flammini, Johan Bollen, Alessandro Vespignani, and other current and past members of the Truthy group at Indiana University  for helpful discussions and contributions to the Truthy Project. We gratefully acknowledge support from the National Science Foundation (grant CCF-1101743), DARPA (grant W911NF-12-1-0037), and the McDonnell Foundation. 

\bibliographystyle{acm-sigchi}
\bibliography{truthydemo}

\section{Figures}

\newpage

  
\balance

\end{document}